\documentclass[12pt]{iopart}

\usepackage{iopams}
\usepackage{setstack}
\usepackage{graphicx}
\usepackage[caption=false]{subfig}

\begin{document}

\title[Disordered ASEPs with bypassing]{Absence of phase coexistence in disordered exclusion processes with bypassing}

\author{J Szavits-Nossan and K Uzelac}
\address{Institute of Physics, Bijeni\v{c}ka cesta 46, HR-10001 Zagreb, Croatia}
\eads{\mailto{juraj@ifs.hr}, \mailto{katarina@ifs.hr}}

\begin{abstract}
Adding quenched disorder to the one-dimensional asymmetric exclusion process is known to always induce phase separation. To test the robustness of this result, we introduce two modifications of the process that allow particles to bypass defect sites. In the first case, particles are allowed to jump $l$ sites ahead with the probability $p_l\sim l^{-(1+\sigma)}$, where $\sigma>1$. By using Monte Carlo simulations and the mean-field approach, we show that phase coexistence may be absent up to enormously large system sizes, e.g. $\mathrm{ln}L \sim 50$, but is present in the thermodynamic limit, as in the short-range case. In the second case, we consider the exclusion process on a quadratic lattice with symmetric and totally asymmetric hopping perpendicular to and along the direction of driving, respectively. We show that in an anisotropic limit of this model a regime may be found where phase coexistence is absent.
\end{abstract}


\maketitle

\section{Introduction}

Disorder usually has a strong impact on critical phenomena, a hallmark example being the influence of spatially uncorrelated quenched disorder on the second-order phase transition, the relevance of which is predicted by the Harris criterion \cite{Harris74}. Such a universal result is, however, generally lacking in systems far from equilibrium. There are few exceptions, and particularly well-studied are driven diffusive systems, in particular the asymmetric simple exclusion process (ASEP), for which even a version of the Harris criterion can be established \cite{Stinchcombe02}. Being exactly solvable \cite{SchutzDomany93}, ASEP has attracted a lot of attention due to its connection with quantum spin chains \cite{GwaSpohn92}, nonequilibrium phase transitions \cite{Krug91} and wide range of applications like surface growth \cite{Krug97}, traffic \cite{Chowdhury00} and biological transport \cite{Chowdhury05}.

ASEP is a simple model describing classical particles hopping in the preferred direction on a discrete lattice and interacting only by the exclusion principle that forbids two particles from occupying the same lattice site (for reviews, see e.g. \cite{Derrida98,Schadscheider10}). Disorder in ASEP is introduced by assigning random (but quenched) hopping rates either to sites or particles. In the simplest case of a single defect particle hopping at rate $r<1$, the exact solution was found \cite{Mallick96,LeePopkovKim97}, with the conclusion that the defect induces phase coexistence through the creation of shocks for all particle densities $\rho<1-r$. A similar conclusion was obtained for a model with an arbitrary number of slow particles solved simultaneously by Evans \cite{Evans96} and Ferrari and Krug \cite{KrugFerrari96}, who observed that the appearance of shocks is determined solely by the slowest particle in the system. On the contrary, ASEP with a single defect site is still an open problem. Although early studies of this problem in various contexts \cite{WolfTang90,JanowskyLebowitz92,TangLyuksyutov93,Schutz93,JanowskyLebowitz94} provided strong arguments that a single defect always induces phase separation (i.e. for all $r<1$), several later studies suggested the existence of a threshold value $r_c<1$ above which the shocks disappear \cite{Ha03,LeeKim09}. On the other hand, no such ``shock-free'' scenario is possible for the full site-wise disorder, as pointed out by Tripathy and Barma \cite{TripathyBarma97,TripathyBarma98} in the particular case of a binomial distribution. The reason behind this is essentially a geometric one, in a sense that disorder creates a long stretch (``bottleneck'') of slow sites that limits the current and ensures the shocks to appear as soon as this limiting value is reached. 

The fact that phase coexistence appears for any distribution of disorder \cite{Krug00} raises the following question: do such disorder-induced large-scale inhomogeneities in density profiles persist if one relaxes the geometrical constraint by allowing the particles to move around defects? The purpose of our  paper is to answer this question by considering two different realizations of bypassing in ASEP with disorder. In the first model we introduce long-range jumps weighted by a probability that decays with distance $l$ as $l^{-(1+\sigma)}$, where $\sigma>1$ to ensure finite current. The choice for this particular model is motivated by the fact that it may lack the phase coexistence in the presence of a single defect \cite{SzavitsUzelac09}. From the technical point of view, in the absence of disorder the model retains most of the characteristics of the short-range ASEP, but is more accurately described by the mean-field approach than in the short-range case \cite{SzavitsUzelac08}. The second model we consider is the short-range ASEP in two dimensions, where disorder is present only in the direction of driving. This model is far from being just academic. It may be considered in a broader context of various transport phenomena in random media, ranging from a fluid flowing through a porous medium \cite{RamaswamyBarma87,BarmaRamaswamy93}, a traffic flow in the presence of obstacles \cite{Saegusa08} to, more recently, a transport of microfluidic droplets through various geometries \cite{Champagne10}.  Common to all these systems is the nontrivial interplay between driving, mutual interactions and the underlying geometry, which altogether may reduce the conductivity, in some cases even completely \cite{RamaswamyBarma87}.

The remainder of the paper is organized as follows. Results for the long-ranged ASEP in one dimension are presented in section 2. The short-range ASEP in two dimensions is introduced in section 3, where we first analyse the current-density relation and then consider the anisotropic limit of hopping rates. A summary of results is presented in section 4.

\section{Bypassing by long-range hopping in one dimension}

\begin{figure}[!hb]
\centering\includegraphics[height=4cm]{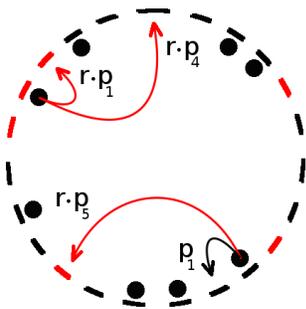}
\caption{A schematic picture of the long-range hopping in disordered TASEP with periodic boundary conditions. Hopping rates are reduced by the factor $r$ whenever the particle jumps from, or to a defect site (red colour). Note that not all the possible moves are sketched, but only those that illustrate the implementation of disorder.}
\label{fig1}
\end{figure}

We consider the model with $N=\rho L$ particles distributed on $L$ sites of a one-dimensional ($1$D)lattice with periodic boundary conditions, where each site is either occupied by a particle ($\tau_n=1$), or is empty ($\tau_n=0$). Dynamics is implemented by randomly choosing one particle during an infinitesimal interval $\rmd t$ and attempting to move it $l$ sites to the right, where $1\leq l<L$ is taken from the probability distribution $p_{l}=l^{-(1+\sigma)}/\zeta_{L-1}(\sigma+1)$, $\zeta_{L-1}(z)$ being the partial sum of the Riemann zeta function. If the target site is empty, the move is accepted; otherwise, it is rejected. Additionally, we introduce $N_{\mathrm{d}}=c L$ ($c$ being concentration) fixed but randomly distributed (defect) sites in a way that the rate of hopping is weighted by the additional factor $r<1$ whenever particle jumps either \emph{to} or \emph{from} a defect site (see \fref{fig1} for illustration). This particular choice of disorder fulfils two purposes: first, the particles can bypass the defect sites by the long-ranged jumps and second, the particle-hole symmetry is preserved. The latter is, in fact, not essential to our conclusions, but makes analysis of results easier.


\subsection{Typical results of Monte Carlo simulations}

\begin{figure}[ht]
\centering
\subfloat{\label{fig2a}\includegraphics[width=6.5cm]{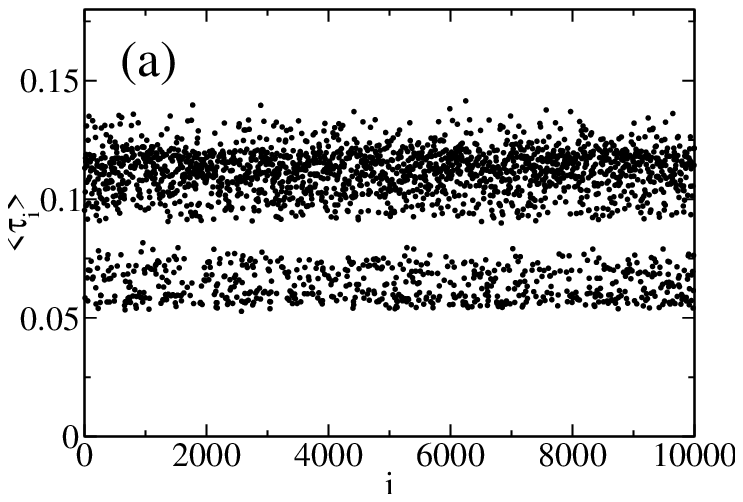}}
\quad                
\subfloat{\label{fig2b}\includegraphics[width=6.5cm]{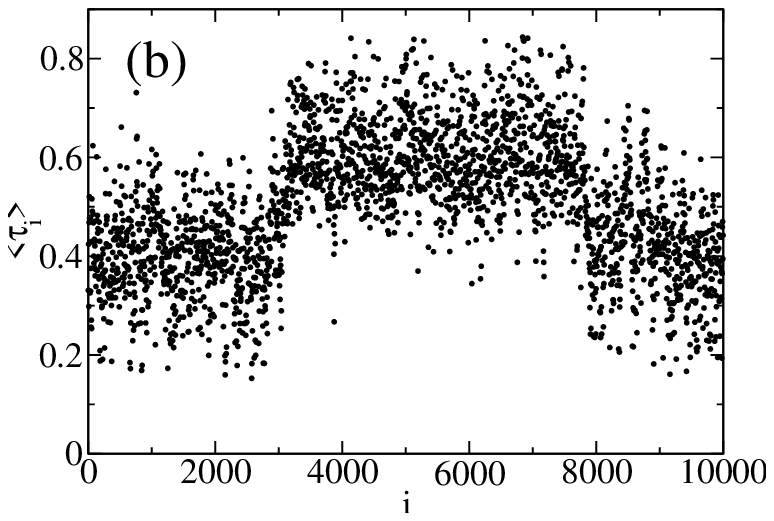}}
\caption{Density profiles for (a) $\rho=0.1$ and (b) $\rho=0.5$ for a system of size $L=10^4$ with $r=0.5$, $c=0.5$ and $\sigma=1.8$ obtained by Monte Carlo simulations ($t=10^7$ MCS/site) for a single 	disorder realization.}
\label{fig2}
\end{figure}

We are interested in the stationary density profiles $\langle\tau_i\rangle$, $i=1,\dots,L$, where $\langle\dots\rangle$ is defined as $\langle\dots\rangle=\sum_{C}(\dots)P(C)$, and $P(C)$ is the stationary solution of the corresponding master equation. Since the stationary solution is not known, the extensive Monte Carlo simulations were performed, representative results of which are presented in figures \ref{fig2a} and \ref{fig2b}. Figure \ref{fig2a} shows the density profile typical of low densities $\rho$, where only the microscopic shocks appear. On the other hand, raising $\rho$ above some threshold value $\rho_{\mathrm{c}}$ induces a macroscopic shock (i.e. of size $\propto L$), as presented in figure \ref{fig2b} (due to the particle-hole symmetry, the same applies for lowering $\rho$ below $1-\rho_{\mathrm{c}}$). 


\begin{figure}[ht]
\centering
\subfloat{\label{fig3a}\includegraphics[width=6.5cm]{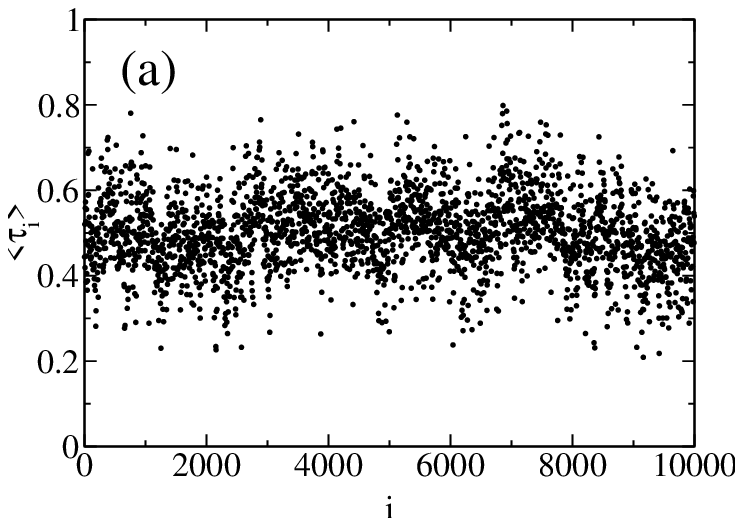}}
\quad                
\subfloat{\label{fig3b}\includegraphics[width=6.5cm]{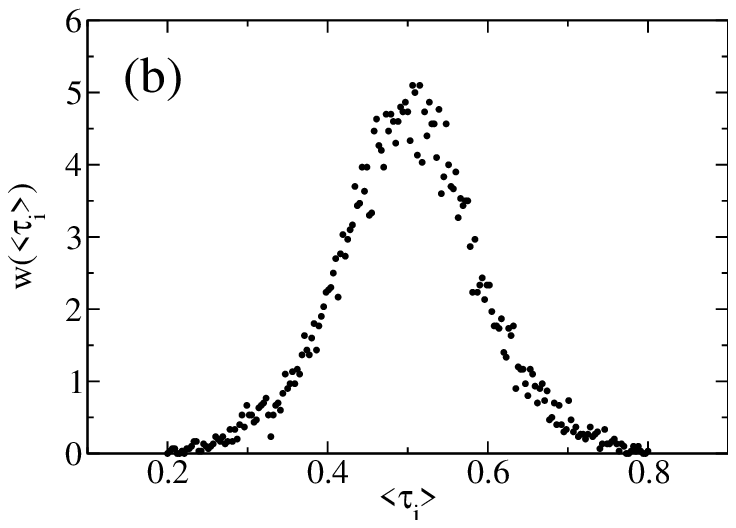}}
\caption{(a) Density profile for a system of size $L=10^4$ with $r=0.5$, $c=0.5$ and $\sigma=1.2$ obtained by Monte Carlo simulations 
($t=10^7$ MCS/site) for a single disorder realization at density $\rho=1/2$ and (b) the corresponding (normalized) histogram of densities 
$\langle\tau_i\rangle$, $i=1\dots,L$ which displays a single maximum around $\rho=1/2$.}
\label{fig3}
\end{figure}

\begin{figure}[ht]
\centering
\includegraphics[width=7.5cm]{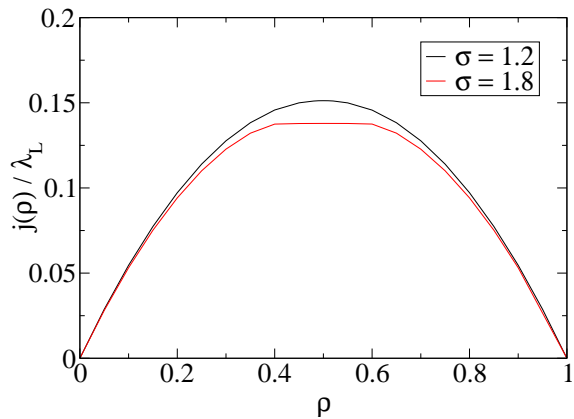}               
\caption{Current-density relation (fundamental diagram) for various $\sigma$ ($L=10^4$, $c=0.5$ and $r=0.5$) obtained by Monte Carlo simulations
($t=10^7$ MCS/site) for a single disorder realization. For better comparison, the current has been scaled with the average hopping length $\lambda_L=\sum_{i}^{L}i\cdot p_i$.}
\label{fig4}
\end{figure}

So far, this is reminiscent of the behaviour found in the short-range case \cite{TripathyBarma98}, except that $\rho_{\mathrm{c}}$ appears to depend strongly on $\sigma$. This fact seems appealing as one may ask whether $\rho_{\mathrm{c}}$ could reach $1/2$ (resulting with absence of macroscopic shocks) by lowering $\sigma>1$ (as it is the case for a single defect \cite{SzavitsUzelac09}). One choice of parameters for which this type of change may be observed in Monte Carlo simulations is presented in figure \ref{fig3a}, along with the corresponding histogram, i.e. frequency distribution of densities $\langle\tau_i\rangle$ (figure \ref{fig3b}). This lack of phase coexistence may also be observed in the current-density relation $j(\rho)$, where the usual plateau is missing (figure \ref{fig4}). 

On the other hand, finding $\rho_{\mathrm{c}}$ for a general choice of parameters $r$, $c$ and $\sigma$ proves to be an elusive task. Even in the short-range case one can only make an estimate, for example, using the so-called fully-segregated model \cite{TripathyBarma98} in which all the defect sites are in a consecutive order forming a large ``bottleneck''. This is the approach we pursue in the following section.


\subsection{Mean-field approach}

In the fully-segregated model of Tripathy and Barma \cite{TripathyBarma98}, all the defects are in a consecutive order forming a large segment (say $y$) of size $N_{\mathrm{d}}=c L$, where the hopping rates are reduced by the factor $r$. Provided this domain is large enough, increasing of density $\rho$ will eventually bring the stationary current $j$ to its maximum value for this domain, $j=r/4$, which corresponds to density $\rho_{\mathrm{y}}=1/2$. Due to the conservation of current, $j$ has to be matched with the current in the remaining segment (say $x$)

\begin{equation}
\label{jx=jy}
\rho_{\mathrm{x}}(1-\rho_{\mathrm{x}})=r\rho_{\mathrm{y}}(1-\rho_{\mathrm{y}}).
\end{equation}

\noindent Together with the conservation of particles, $\rho_{\mathrm{x}}(1-c)+c\rho_{\mathrm{y}}=\rho$, this gives the following (mean-field) estimate of $\rho_{\mathrm{c}}$

\begin{equation}
\label{rhoc}
\rho_{\mathrm{c}}=\frac{1-(1-c)\sqrt{1-r}}{2}.
\end{equation}

\noindent Later, Krug \cite{Krug00} showed that (\ref{rhoc}) is actually the upper bound on the exact value of $\rho_{\mathrm{c}}$ in the disordered model, while the lower bound is given by

\begin{equation}
\label{rhoc_lower}
\rho_{\mathrm{c}}\geq\frac{1-\sqrt{1-r}}{2}.
\end{equation}

\noindent Note that these two bounds meet in the limit of infinitesimal concentration $c\rightarrow 0$ (not to be confused with a single defect!) giving $\rho_{\mathrm{c}}=(1-\sqrt{1-r})/2\neq 1/2$. 

\begin{figure}[!hb]
\centering
\subfloat{\label{fig5a}\includegraphics[width=7cm]{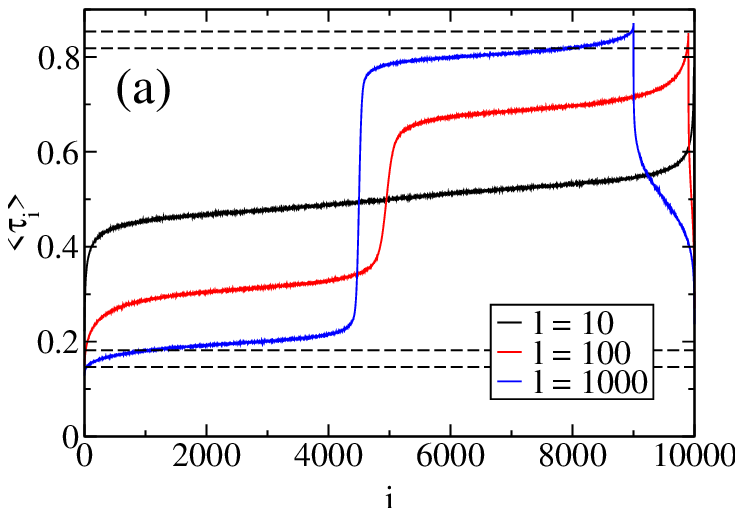}}
\quad                   
\subfloat{\label{fig5b}\includegraphics[width=7cm]{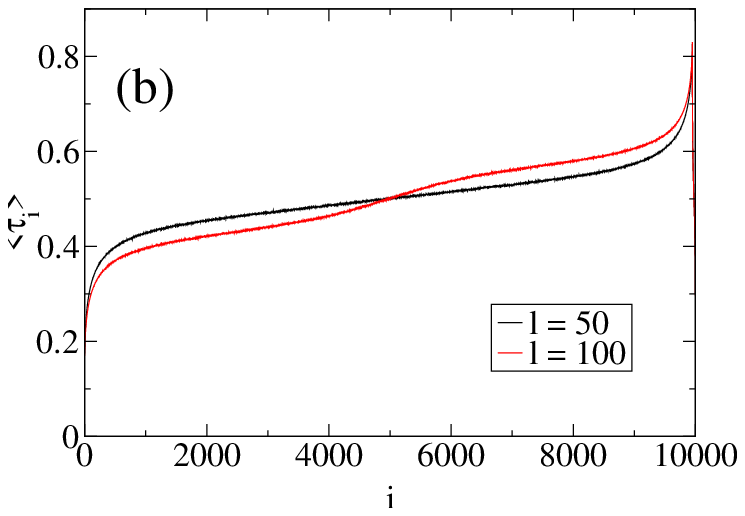}}
\caption{Density profiles in the segregated model for various $l$ ($L=10^4$, $r=0.5$) obtained by Monte Carlo simulations 
($t=10^7$ MCS/site) for a single disorder realization at density $\rho=1/2$ for (a) $\sigma=1.2$ and (b) $\sigma=1.05$. In (a), dashed lines
represent asymptotic values of $\rho_{\mathrm{c}}$ (and $1-\rho_{\mathrm{c}}$) according to bounds (\ref{rhoc}) and (\ref{rhoc_lower}) with $c=l/L$ and $l=1000$. In
(b), onset of shock creation is captured for $l$ that corresponds to system sizes unreachable in Monte Carlo simulations with full disorder.}
\label{fig5}
\end{figure}

For these results to be applied to the original, fully disordered system, one argues that the largest ``bottleneck'' has diverging length $l\sim \mathrm{ln} L$ in the limit $L\rightarrow\infty$ \cite{TripathyBarma98}, so that the current eventually establishes its asymptotic value $j_{\infty}=r\rho_{\mathrm{y}}(1-\rho_{\mathrm{y}})$. The same is true in the long-range case, except that the current $j_{\infty}$ gains an additional factor $\lambda=\sum_{i}i\cdot p_i$, $j_{\infty}=\lambda\cdot r\rho_{\mathrm{y}}(1-\rho_{\mathrm{y}})$. Since this has no effect on (\ref{jx=jy}), one expects the same $\rho_{\mathrm{c}}$ as in (\ref{rhoc}). This means that the absence of a macroscopic shock in figure \ref{fig3a} should be a {\it finite-size effect} and should be ruled out in the infinite system. However, as we shall see later, to observe it in Monte Carlo simulations would require enormous system sizes $L$ (e.g. $\mathrm{ln}L\approx 50$, see \ref{fig5b}). [It is worth mentioning that such a slow approach to the thermodynamic limit was found in some other processes in driven diffusive systems \cite{KafriLevine02}.] In what follows we therefore adopt the segregated model of Tripathy and Barma, but with $l$ as a free parameter. i.e. not related to $L$.


The results obtained by increasing $l$ at two different values of $\sigma$, $1.2$ and $1.05$, are presented in figures \ref{fig5a} and \ref{fig5b}, respectively. The density profiles are similar to those in the presence of a single defect, where shocks disappear with increasing $r$ (short-range model) or with decreasing $\sigma$ (long-range model). The main difference is that in the single defect case, the system is able to attain the same value of the maximum current as in the {\it pure} model, which is done by building long-range correlations in the density profile. On the other hand, in the segregated or fully disordered models, this is not possible as the largest ``bottleneck'', whose length diverges with $L$, will force $j$ to a value less than in the pure case, provided $l$ is large enough. How large $l$ is needed to establish the asymptotic regime is not an easy problem as it depends on the contribution to the current coming from the site-dependent corrections to $\rho_{\mathrm{x}}$ and $\rho_{\mathrm{y}}$ (for example, in figure \ref{fig5b}, $\mathrm{ln}L=50$ is still not enough to observe shock). Instead of calculating these corrections explicitly, in what follows we estimate their leading contribution to the total current as a function of $l$.  

\indent Let us enumerate lattice sites so that ``bottleneck'' occupies sites $i=L-l+1,\dots,L$. In the long-range model, the current $j_i$ that satisfies local conservation law,

\begin{equation}
\frac{d}{dt}\langle\tau_i\rangle(t)=j_{i-1}-j_{i},
\end{equation}

\noindent is defined as the total current of all particles jumping from and over the site $i$,

\begin{equation}
\label{jot}
j_i=\sum_{m=0}^{L-1}\sum_{m+n<L}p_{m+n}\langle\tau_{i-m}(1-\tau_{i+n})\rangle\delta_{i-m,i+n}^{r},
\end{equation}

\noindent where disorder is introduced through $\delta_{k,l}^{r}$ equal to $r$ or $1$ depending on whether a pair of sites $k,l$ contains at least one defect site or not, respectively. Note also that, due to the periodic boundary conditions, $\tau_{j\pm L}=\tau_j$, $j=1,\dots,L$. To calculate the current, we choose the site in the middle of the ``bottleneck'' ($i=L-l/2+1$) and apply the mean-field approximation, $\langle\tau_i\tau_j\rangle\approx\langle\tau_i\rangle\langle\tau_j\rangle$.  Assuming the following density profile 

\begin{equation}
\label{smf}
\langle\tau_i\rangle=\cases{\rho_{i}^{\mathrm{x}}, &$1\leq i\leq L-l$\\ \rho_{i}^{\mathrm{y}} ,& $L-l+1\leq i\leq L$,}
\end{equation}

\noindent gives four contributions to $j$ ($j_{\mathrm{xx}}$, $j_{\mathrm{xy}}$, $j_{\mathrm{yx}}$ and $j_{\mathrm{yy}}$) coming from the exchange of particles between $x$ and $y$ segments. Since we are interested only in how fast $j$ converges with $l$, we can ignore $\rho_{i}^{\mathrm{x,y}}$ as they are bounded between $0$ and $1$. It is then straightforward to show that
\numparts
\begin{eqnarray}
j_{\mathrm{xx}} &<& r(1\cdot p_1+2\cdot p_2+\dots+(l/2)p_{l/2}) \sim\Or(1)\label{jxx}\\
j_{\mathrm{xy}},j_{\mathrm{yx}} &<& r(1\cdot p_{l/2+1}+2\cdot p_{l/2+2}+\dots+(l/2)p_{l}) \sim\Or(l^{-(\sigma-1)})\label{jxy} \\
j_{\mathrm{yy}} &< & 1\cdot p_{l+1}+2\cdot p_{l+2}+\dots+lp_{l}\sim\Or(l^{-(\sigma-1)}),\label{jyy}
\end{eqnarray}
\endnumparts

\noindent so that contributions (\ref{jxy}) and (\ref{jyy}) decay very slowly when $\sigma$ is close to $1$. 

To conclude, site-wise disorder in one dimension seems too restrictive, even if defects can be bypassed by the long-ranged jumps. As concluded by Tripathy and Barma, the reason for this is that disorder in one dimension generates ``bottleneck'' of slow sites that diverges in the limit $L\rightarrow\infty$. However, such geometric constraint is not to be expected in higher dimensions, which we consider in the rest of this paper.

\section{Bypassing by transverse hopping in two dimensions}

We consider $N=\rho L_{\mathrm{x}} L_{\mathrm{y}}$ particles distributed on a two-dimensional ($2$D) quadratic lattice consisting of $L_{\mathrm{x}}\times L_{\mathrm{y}}$ sites, where each site holds at most one particle at a time ($\tau_{ij}\in\{0,1\}$). At any given moment, a randomly chosen particle attempts to move to the nearest site either in the direction of driving with the probability $p_{\parallel}=p_{\mathrm{x}}$ or in the perpendicular direction with the probability $p_{\perp}=p_{\mathrm{y}}$ so that $p_{\mathrm{x}}+2 p_{\mathrm{y}}=1$. If the target site is empty, the move is accepted; otherwise, it is rejected. In other words, the exclusion process is totally asymmetric in the $\hat{x}$ direction and symmetric in the $\hat{y}$ direction with periodic boundary conditions assumed in both directions. To maximally ease bypassing, disorder is introduced by randomly choosing $cL_{\mathrm{x}} L_{\mathrm{y}}$ sites from where hopping in the direction of driving is forbidden, which could be imagined as if a ``bond'' between the two adjacent sites in $\hat{x}$ direction was broken. In other words, a particle at a defect site either moves transverse to the driving with probability $p_{\mathrm{y}}$ (provided the target site is empty) or waits with the probability $p_{\mathrm{x}}$ until the next attempt.

Two-dimensional exclusion process has been studied previously in various contexts. In \cite{RamaswamyBarma87,BarmaRamaswamy93}, Ramaswamy and Barma have studied a partially asymmetric exclusion process with $p_{\mathrm{up}}=p_{\mathrm{right}}=w\cdot(1+g)$ and $p_{\mathrm{left}}=p_{\mathrm{down}}=w\cdot(1-g)$, where disorder is introduced by breaking ``bonds'' in both $\hat{x}$ and $\hat{y}$ directions with probability $c$. The underlying network created in that way is somewhat different from ours, because it creates backbends along which current flows against the driving field. Even closer to our problem is the work of Saegusa \emph{et al} \cite{Saegusa08}, who considered flow of particles in the multi-lane TASEP ($L_{\mathrm{y}}\ll L_{\mathrm{x}}$) with fixed obstacles, where the usual plateau in the current-density relation was observed. Alexander and Lebowitz \cite{AlexanderLebowitz94} considered the motion of a rod immersed in a fluid of interacting particles described by the symmetric simple exclusion process and observed a macroscopic region of low density behind the rod. Finally, large-scale inhomogeneities in two-dimensional driven diffusive systems, although not induced by disorder but akin to phase coexistence, have been investigated in many works, e.g. by Schmittmann \emph{et al} \cite{SchmittmannHwangZia92}. 

\subsection{Typical results of Monte Carlo simulations} 

\subsubsection{Current-density relation}

\begin{figure}[!ht]
\centering
\includegraphics[width=7.5cm]{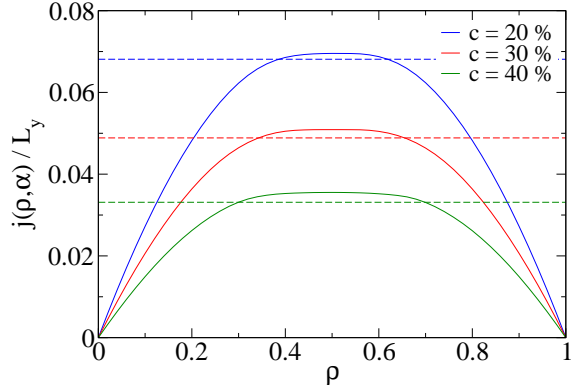}               
\caption{Current-density relation (fundamental diagram) for various concentrations $c$ on $200\times 200$ lattice with $p_{\mathrm{x}}=2 p_{\mathrm{y}}=1/2$, each obtained by Monte Carlo simulations ($t=10^6$ MCS/site) for a single disorder realization. Dashed lines denote the best estimate of the maximum current given by the expression (\ref{jotx_approx}).}
\label{fig6}
\end{figure}

\indent Our first task is to understand how the presence of disorder affects the stationary current in the $\hat{x}$ direction, which is defined as

\begin{equation}
j_{\mathrm{x}}(\rho,\alpha)/L_{\mathrm{y}}=\frac{1}{L_{\mathrm{y}}}\sum_{j=1}^{L_{\mathrm{y}}}p_{\mathrm{x}}\langle\tau_{ij}(1-\tau_{i+1,j})\rangle\cdot\omega_{ij}(\alpha), \quad i=1,\dots,L_{\mathrm{x}},
\label{jotx}
\end{equation}

\noindent where, for a given realization of disorder $\alpha$, $\omega_{ij}(\alpha)=1$ if the bond that connects sites at $(i,j)$ and $(i+1,j)$ is present, and $\omega_{ij}(\alpha)=0$ otherwise. Figure \ref{fig6} shows $j(\rho)$ obtained by Monte Carlo simulations for various concentrations $c$ on $200\times 200$ lattice with $p_{\mathrm{x}}=2 p_{\mathrm{y}}=1/2$. Compared to the pure case where $j_{\mathrm{x}}(\rho)/L_{\mathrm{y}}=p_{\mathrm{x}}\rho(1-\rho)$, we find that, before reaching its maximum, the current can be quite well fitted to the parabolic shape, $j(\rho)\propto\rho(1-\rho)$, but the factor seems nontrivial and decreases with increasing concentration $c$. More importantly, one observes a plateau around $\rho=1/2$, but its boundaries are less sharp compared to the one-dimensional model. 

A naive estimate of height of the plateau can be given by the average number of regular bonds per column times $p_{\mathrm{x}}/4$, $(1-c)p_{\mathrm{x}}/4$, which is too high as seen from the third column in table \ref{tab}. This estimate can be improved by looking at the column with the smallest number of regular bonds, which can be either found directly in the particular realization of disorder or calculated using the extreme value theory \cite{extremevalue}. Since we could not find any reference in literature to point to this problem explicitly, we present details of this calculation in the Appendix and state here only the final result
\begin{eqnarray}
\omega^{*}(\alpha)\equiv\underset{i}{min}\left\{\sum_{j=1}^{L_{\mathrm{y}}}\omega_{ij}(\alpha)\right\}&=& L_{\mathrm{y}}-\underset{i}{max}\left\{\sum_{j=1}^{L_{\mathrm{y}}}[1-\omega_{ij}(\alpha)]\right\}\approx\nonumber \\ &\approx& L_{\mathrm{y}}-a_{L_{\mathrm{x}}}(c,L_{\mathrm{y}})\gamma-b_{L_{\mathrm{x}}}(c,L_{\mathrm{y}})\equiv{\overline{\omega}}^{*},
\label{ab}
\end{eqnarray}


\noindent where $\gamma=0.5772\dots$ is Euler-Mascheroni constant and $a_{L_{\mathrm{x}}}(c)$ and $b_{L_{\mathrm{x}}}(c)$ are given by the expressions (\ref{a}) and (\ref{b}), respectively, in the Appendix. [Note that ${\overline{\omega}}^{*}$ in (\ref{ab}) is independent of $\alpha$ since it is obtained by averaging over all disorder configurations with concentration $c$. However, since $L_{\mathrm{x}}$ is generally large, this value is close to $\omega^{*}(\alpha)$ obtained by counting regular bonds explicitly from a particular realization of disorder $\alpha$ (as seen from the fourth and the fifth columns in table \ref{tab}).] Although the analytical expression (\ref{ab}) is closer to Monte Carlo data than $(1-c)p_{\mathrm{x}}/4$, it poorly describes the maximum current overall, especially at higher concentrations $c$. The best estimate, as displayed in the sixth column of table \ref{tab}, is obtained if one recognizes that the largest contribution to the current will come from those sites (within a column) that have both inward and outward bonds, which gives
\begin{eqnarray}
\fl\qquad \underset{\rho}{max}\{j_{x}(\rho,\alpha)\}/L_{\mathrm{y}}&\approx& \frac{1}{L_{\mathrm{y}}}\underset{i}{min}\left\{\sum_{j=1}^{L_{\mathrm{y}}}\omega_{i-1,j}(\alpha)\omega_{ij}(\alpha)\right\}\cdot\frac{p_{\mathrm{x}}}{4}\approx\nonumber \\
&\approx& [L_{\mathrm{y}}-a_{L_{\mathrm{x}}}(2c-c^2,L_{\mathrm{y}})\gamma-b_{L_{\mathrm{x}}}(2c-c^2,L_{\mathrm{y}})]\cdot p_{\mathrm{x}}/(4L_{\mathrm{y}}),
\label{jotx_approx}
\end{eqnarray}

\noindent where $2c-c^2=1-(1-c)^2$ is the probability of not having both inward and outward bond at the same site. It should be noted that the usefulness of this expression depends on the value of $p_{\mathrm{x}}$: in the strongly anisotropic limit where $p_{\mathrm{x}}\ll 2p_{\mathrm{y}}$, there will be an increasing contribution to the current coming from the particles entering the column at one site and then leaving it from another. Details of this analysis should be published elsewhere.

\begin{table}[!ht]
\caption{\label{tab}Values of maximum current for various concentration $c$ obtained from Monte Carlo simulations ($L_{\mathrm{x}}=L_{\mathrm{y}}=200$, $p_{\mathrm{x}}=2p_{\mathrm{x}}=1/2$, $t=10^6$ MCS/site) and compared to a naive guess $(1-c)p_{\mathrm{x}}/4$, the expression $\omega^{*}(\alpha)p_{\mathrm{x}}/(4L_{\mathrm{y}})$ obtained by counting bonds explicitly from $\alpha$, the expression ${\overline{\omega}}^{*}p_{\mathrm{x}}/(4L_{\mathrm{y}})$ calculated from the extreme value theory and to the best estimate given by the expression (\ref{jotx_approx}).}
\begin{indented}
\item[]\begin{tabular}{@{}llllll}
\br
$c$ & Monte Carlo & $(1-c)p_{\mathrm{x}}/4$ & $\omega^{*}(\alpha)\cdot p_{\mathrm{x}}/(4L_{\mathrm{y}})$ & ${\overline{\omega}}^{*}\cdot p_{\mathrm{x}}/(4L_{\mathrm{y}}) $ & expression (\ref{jotx_approx})\\
\mr
$0.1$&0.09255(5)&0.1125&0.10507&0.10313&0.09154 \\
$0.2$&0.06953(7)&0.1000&0.09063&0.09010&0.06812 \\
$0.3$&0.05089(2)&0.0875&0.07688&0.07615&0.04887 \\
$0.4$&0.03553(5)&0.0750&0.06000&0.06287&0.03312 \\
\br
\end{tabular}
\end{indented}
\end{table}

\subsubsection{Density profiles}

\begin{figure}[!ht]
\centering
\includegraphics[width=10cm]{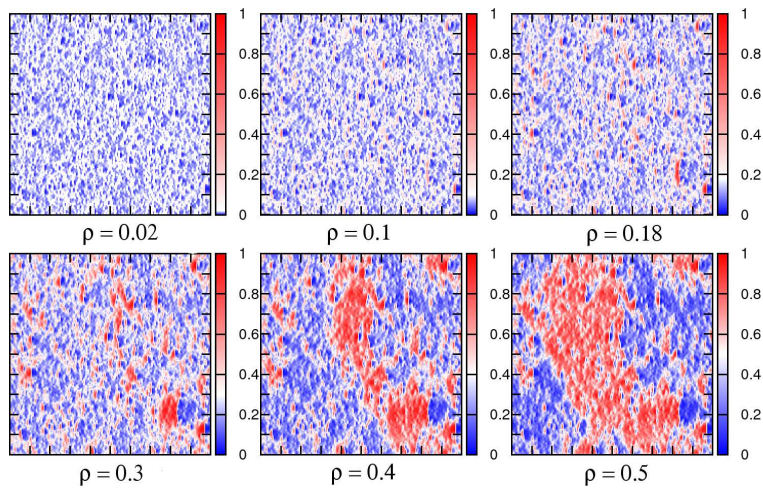}               
\caption{Density profiles for various $\rho$ at $c=0.4$ obtained by Monte Carlo simulations ($L_{\mathrm{x}}=L_{\mathrm{y}}=200$, $p_{\mathrm{x}}=2p_{\mathrm{x}}=1/2$, $t=10^6$ MCS/site) for a single disorder realization, represented by the shades of blue ($\langle\tau_{ij}\rangle=0$), white ($\langle\tau_{ij}\rangle=\rho$) and red ($\langle\tau_{ij}\rangle=1$).}
\label{fig7}
\end{figure}

Having found some basic characteristics of the current-density relation for typical values of $p_{\mathrm{x}}$ and $p_{\mathrm{y}}$, we now turn to the investigation of stationary density profiles. Figure \ref{fig7} shows the density profiles obtained by Monte Carlo simulations for various $\rho$, represented by the shades of blue ($\langle\tau_{ij}\rangle=0$), white ($\langle\tau_{ij}\rangle=\rho$) and red ($\langle\tau_{ij}\rangle=1$).


\begin{figure}[!ht]
\centering
\subfloat{\label{fig8a}\includegraphics[height=5cm]{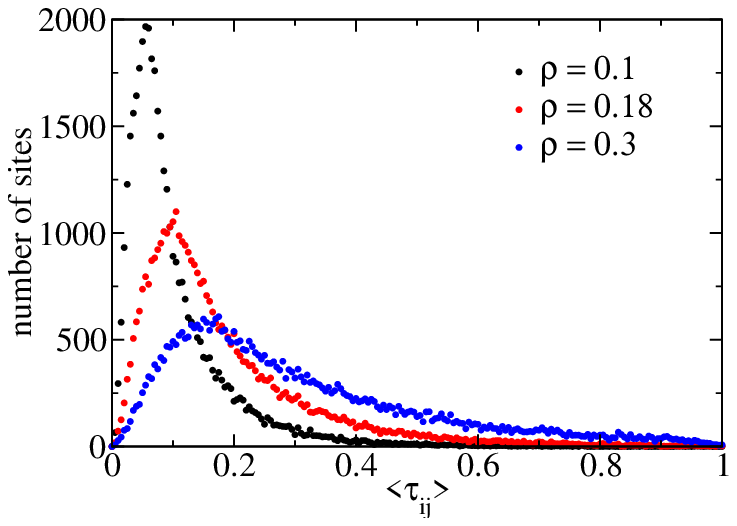}}
\quad                   
\subfloat{\label{fig8b}\includegraphics[height=5cm]{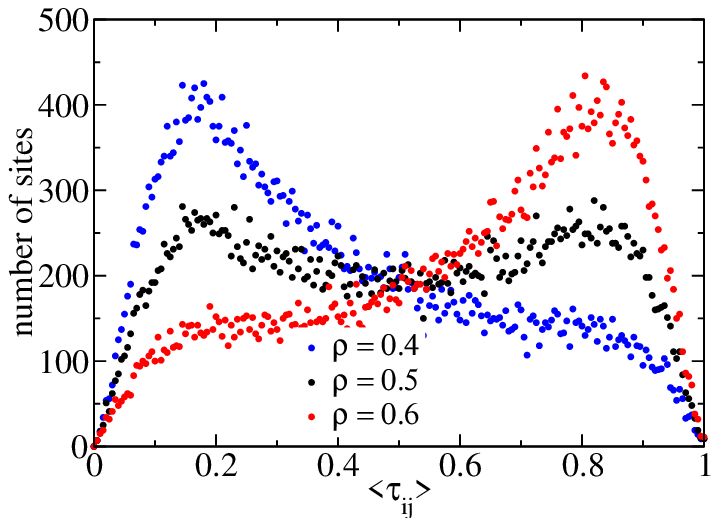}}
\caption{Histogram of density profiles obtained by Monte Carlo simulations for a single disorder configuration at $c=0.4$ ($L_{\mathrm{x}}=L_{\mathrm{y}}=200$, $p_{\mathrm{x}}=2p_{\mathrm{y}}=1/2$, $t=10^6$ MCS/site) displaying (a) single maximum at lower $\rho$ and (b) two maxima at intermediate $\rho$.}
\label{fig8}
\end{figure}

Inspecting profiles for increasing $\rho$, one clearly sees how local inhomogeneities present at lower $\rho$, start to grow until two large-scale phases occur for $\rho$ around $0.4$. This is even better seen from the corresponding histograms, which display either a single maximum at lower/higher densities (figure \ref{fig8a}) or two maxima at intermediate densities (figure \ref{fig8b}). [It is also interesting to observe that even when $\rho$ is small, density distribution is still not peaked around $\rho$.] Whether this is a true transition or just a finite-size effect seems difficult to determine using only numerical approaches, as they are restricted to not too large system sizes. We therefore tackle the problem from a different angle by looking at the strongly anisotropic limit $p_{\mathrm{x}}\rightarrow 0$, where a rather simple result emerges that predicts no phase coexistence, no matter what $c$ is.


\subsection{Mean-field approach in the limit $p_{\mathrm{x}}\rightarrow 0$}

In the limit $p_{\mathrm{x}}\rightarrow 0$, the jumps in the $\hat{x}$ direction are rare compared to the $\pm \hat{y}$ direction and $n_i=\sum_{j}\tau_{ij}$ may be taken as a slow variable whose evolution in time is described by the following master equation:

\begin{equation}
\fl \qquad\frac{\rmd}{\rmd t}P(C,t)=\sum_{i}\left[W(C_{+}^{i,i+1}\rightarrow C)P(C_{+}^{i,i+1},t)-W(C\rightarrow C_{-}^{i,i+1})P(C,t)\right],
\label{masterbox}
\end{equation}

\noindent where $C=\{n_i\vert i=1,\dots,L_{\mathrm{x}}\}$ and $C_{\pm}^{i,i+1}=\{n_1,\dots,n_i\pm 1,n_{i+1}\mp 1,\dots,n_{L_{\mathrm{x}}}\}$. The next question is how to properly describe transition rates $W$. Since jumps between columns are rare, we may assume that particles have enough time to distribute uniformly within each column, so that each site within a column $i$ has a probability $n_i/L_{\mathrm{y}}$ of holding a particle. Under this assumption, the transition rates read
\begin{eqnarray}
W(C\rightarrow C_{-}^{i,i+1})&=&p_{\mathrm{x}}\cdot\left(\sum_{j=1}^{L_{\mathrm{y}}}\omega_{ij}\right)\cdot\frac{n_{i}}{L_{\mathrm{y}}}\cdot\left(1-\frac{n_{i+1}}{L_{\mathrm{y}}}\right),\\
W(C_{+}^{i,i+1}\rightarrow C)&=&p_{\mathrm{x}}\cdot\left(\sum_{j=1}^{L_{\mathrm{y}}}\omega_{ij}\right)\cdot\frac{n_{i}+1}{L_{\mathrm{y}}}\cdot\left(1-\frac{n_{i+1}-1}{L_{\mathrm{y}}}\right),
\label{wbox}
\end{eqnarray}
\noindent where $\omega_i\equiv\sum_{j=1}^{L_{\mathrm{y}}}\omega_{ij}$. We have thus reduced the starting two-dimensional problem to the one-dimensional in which disorder is associated with each site through $\omega_i$. In this process, known as the misanthrope process (MP), the hopping rates depend on the position of the particular site as well as on the number of particles both at the departure site and at the target site. 

Let us write down the lattice equation for the average number of particles at site $i$

\begin{equation}
\fl \qquad\frac{\rmd}{\rmd t}\langle n_i(t)\rangle=p_{\mathrm{x}}\cdot\omega_{i-1}\cdot\left\langle\frac{n_{i-1}}{L_{\mathrm{y}}}\cdot\left(1-\frac{n_{i}}{L_{\mathrm{y}}}\right)\right\rangle-p_{\mathrm{x}}\cdot\omega_{i}\cdot\left\langle\frac{n_{i}}{L_{\mathrm{y}}}\cdot\left(1-\frac{n_{i+1}}{L_{\mathrm{y}}}\right)\right\rangle.
\end{equation}

\noindent Note that this equation transforms into mean-field equation for the one-dimensional TASEP if we further assume $\langle n_i n_{i+1}\rangle\approx\langle n_i\rangle\langle n_{i+1}\rangle$ and use $\langle n_i\rangle\rightarrow \langle n_i\rangle/L_{\mathrm{y}}\equiv\rho_i$ and $\omega_i\rightarrow \omega_i/L_{\mathrm{y}}\equiv r_i$, which gives

\begin{equation}
\frac{\rmd\rho_i}{\rmd t}=p_{\mathrm{x}}\cdot r_{i-1}\rho_{i-1}(1-\rho_{i})-p_{\mathrm{x}}\cdot r_{i}\rho_{i}(1-\rho_{i+1}),
\end{equation}

\noindent where the $r_i$ are taken from the binomial distribution

\begin{equation}
P(n=r_i\cdot L_{\mathrm{y}})=\left({L_{\mathrm{y}} \atop n}\right)(1-c)^nc^{L_{\mathrm{y}}-n}
\label{binomial}
\end{equation}

\noindent with the mean and the variance given by $\{r_i\}=1-c$ and $\{r_{i}^{2}\}-\{r_i\}^2=c(1-c)/L_{\mathrm{y}}$, respectively. 

\begin{figure}[!ht]
\centering
\subfloat{\label{fig9a}\includegraphics[height=5cm]{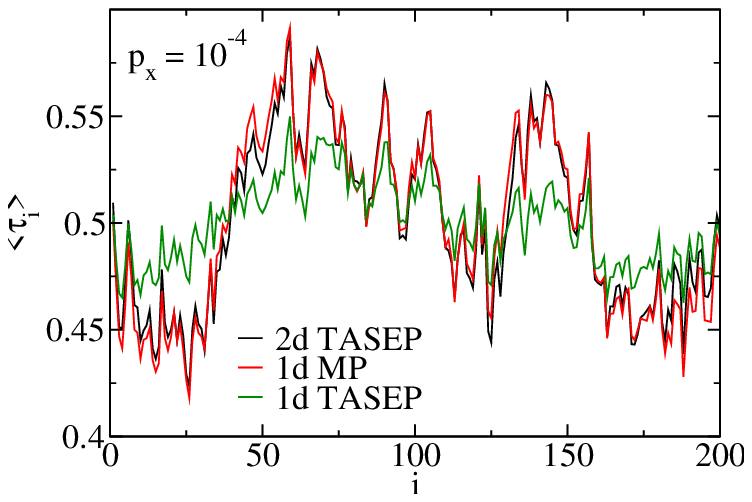}}
\quad                  
\subfloat{\label{fig9b}\includegraphics[height=5cm]{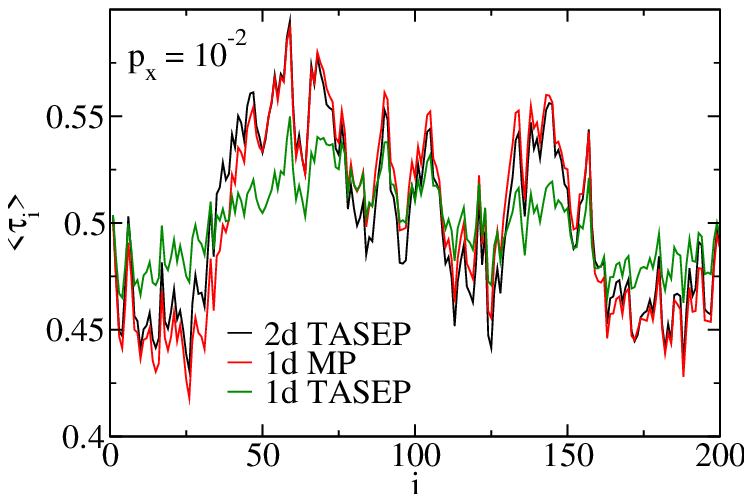}}
\caption{Column-averaged density profiles ($\langle\tau_i\rangle\equiv\sum_{j=1}^{L_y}\langle\tau_{ij}\rangle/L_y$) obtained from Monte Carlo simulations of $2d$ TASEP for (a) $p_{\mathrm{x}}=10^{-4}$ and (b) $p_{\mathrm{x}}=10^{-2}$  ($L_{\mathrm{x}}=L_{\mathrm{y}}=200$, $c=0.2$, $\rho=1/2$, $t=10^6$ MCS/site) compared to density profiles obtained from the equivalent $1d$ misanthrope process (MP) described by (\ref{masterbox})-(\ref{wbox}) and from the $1d$ TASEP.}
\label{fig9}
\end{figure}

To check how well the process defined by (\ref{masterbox})-(\ref{wbox}) describes the original data from the two-dimensional TASEP in the limit $p_{\mathrm{x}}\rightarrow 0$, we performed Monte Carlo simulations with the $r_i$ derived from the same realization of disorder as in $2d$ TASEP and obtained stationary densities $\langle n_i\rangle$. The results are presented in figures \ref{fig9a} and \ref{fig9b} for $p_{\mathrm{x}}=10^{-4}$ and $p_{\mathrm{x}}=10^{-2}$, respectively, together with the data from the corresponding $2d$ and $1d$ TASEP. From these figures it is clear that the one-dimensional misanthrope process (MP) defined by (\ref{masterbox}) and (\ref{wbox}) describes $2d$ data very well for several orders in $p_{\mathrm{x}}$, while the data from $1d$ TASEP follow all the ``peaks'' but show generally less deviation from $1/2$. As we raise $p_{\mathrm{x}}$ further, it is evident that the picture of independent columns with uniformly distributed particles has to be abandoned. A crude estimate when this happens is when the relaxation time of fluctuations within a column, $\tau_{\mathrm{w}}\sim L_{y}^{z}$, is comparable to the average time that a particle spends in one column, $\tau_{\mathrm{c}}\sim 1/p_{\mathrm{x}}$. Since for the symmetric simple exclusion process $z=2$, this gives $p_{\mathrm{x}}\sim L^{-2}$, which seems too small compared to the observed data from Monte Carlo simulations. The precise $p_{\mathrm{x}}$ when this happens is thus an open problem and further work in this direction is needed.


The ``reduction'' of the original two-dimensional problem to the one-dimensional allows us to invoke the same geometric argument as in $1d$ when discussing phase coexistence. However, a striking difference between distribution (\ref{binomial}) of $r_i$'s and the ones previously considered in the 1d disordered TASEP \cite{TripathyBarma98,HarrisStinchcombe04} is in the variance. While in previous studies the variance always remained finite, here it scales as $1/L_{\mathrm{y}}$ meaning that in the limit $L_{\mathrm{y}}\rightarrow\infty$ it becomes improbable to observe an infinitely large domain of sites with, say, all $r_i$ less than some fixed value $r_0$ below average. To see this, let us recall the result obtained by the extreme value theory which gives the average length of the longest sequence of consecutive heads in $N$ coin tosses with $P(head)=p$ \cite{GordonSchillingWaterman86},

\begin{equation}
\{l_{\mathrm{max}}\}=\frac{\gamma+\mathrm{ln}[N(1-p)]}{\mathrm{ln}(1/p)}-\frac{1}{2},
\label{lmax}
\end{equation}

\noindent where $\{\dots\}$ denotes the average over all possible outcomes in $N$ coin tosses. As mentioned earlier, this gives $\{l_{\mathrm{max}}\}\sim \mathrm{ln}L$ when applied to the one-dimensional TASEP of size $N$ with a binomial distribution of defects, where $p$ is the concentration of defect sites. If instead of a binomial distribution one considers (\ref{binomial}), the ``bottleneck'' may be defined as a domain of consecutive sites with all $r_i$'s less then some fixed value $r_0<1-c$, $p=P(x<r_0)$, as it was done by Krug in \cite{Krug00}. Approximating the distribution (\ref{binomial}) with the normal distribution with the mean and the variance given by $\mu=1-c$ and $\sigma^2=c(1-c)/L_{\mathrm{y}}$, respectively, gives 

\begin{equation}
p=P(x<r_0)=\frac{1}{2}\left[1+\mathrm{erf}\left(\frac{r_0-\mu}{\sqrt{2\sigma^2}}\right)\right].
\label{p}
\end{equation}

\noindent Now, if $L_{\mathrm{y}}$ is finite, both $\sigma$ and $p$ are finite, which gives $\{l_{\mathrm{max}}\}\sim \mathrm{ln}L_{\mathrm{x}}$, i.e. an infinitely long ``bottleneck'' in the limit $L_{\mathrm{x}}\rightarrow\infty$. This result applies to the multi-lane TASEP with a finite number of lanes, for which this type of disorder should always induce phase separation. However, in the true $2d$ system of linear size $L$, where $L_{\mathrm{x}}\sim L_{\mathrm{y}}\sim L$, the variance $\sigma\rightarrow 0$ in the limit $L_{\mathrm{y}}\rightarrow\infty$. The error function $\mathrm{erf}(x)$ in (\ref{p}) can be then expanded around $\infty$,

\begin{equation}
\mathrm{erf}(x)=1-\frac{\rme^{-x^2}}{\sqrt{\pi}x}\left[1-\frac{1}{2x^2}+\dots\right],\quad x\rightarrow\infty
\label{erf}
\end{equation}

\noindent If, say, $r_0<\mu=1-c$\footnote{Similarly, one can set $r_0>\mu=1-c$ and then use the expansion (\ref{erf}) around $\infty$.} we may expand the error function around $-\infty$ using the fact that it is an odd function, $\mathrm{erf}(x)=-\mathrm{erf}(-x)$, which after some algebraic manipulation gives

\begin{equation}
\mathrm{ln}(1/p)=\mathrm{ln}\left[\frac{2\sqrt{\pi}\vert r_0-1+c\vert}{\sqrt{c(1-c)}}\right]+\frac{1}{2}\mathrm{ln}L_{\mathrm{y}}+2\frac{(r_0-1+c)^2}{c(1-c)}L_{\mathrm{y}}.
\end{equation}

\noindent In the limit $L_{\mathrm{y}}\rightarrow\infty$, the denominator in (\ref{lmax}) diverges faster that numerator making $\{l_{\mathrm{max}}\}$ vanish with increasing $L_{\mathrm{y}}$. In other words, in the limit $p_{\mathrm{x}}\rightarrow 0$ it is as if ``disorder averages itself'' and the argument in favour of phase coexistence due to the large ``bottleneck'' fails. 

\begin{figure}[!ht]
\centering
\subfloat{\label{fig10a}\includegraphics[height=5cm]{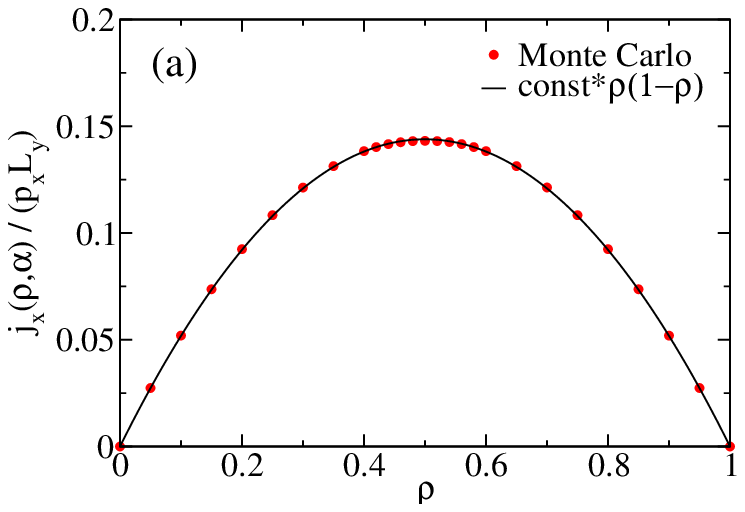}}    
\quad            
\subfloat{\label{fig10b}\includegraphics[height=5cm]{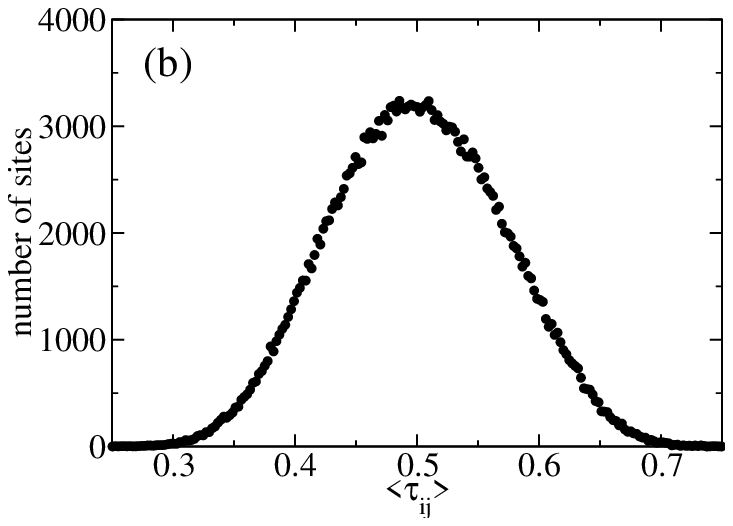}}
\caption{(a) Current-density diagram obtained by Monte Carlo simulations of $2d$ TASEP for a single disorder configuration at $c=0.4$ ($p_{\mathrm{x}}=10^{-2}$, $L_{\mathrm{x}}=L_{\mathrm{y}}=1000$ and $t=10^5$ MCS/site). Solid line is the best fit to the expression $j(\rho)=const.\times\rho(1-\rho)$ with $const.\approx 0.5756$ close to $1-c=0.6$. (b) The corresponding histogram of density profile for $\rho=1/2$.}
\label{fig10}
\end{figure}

To check this in the original, $2d$ model, we present for illustration Monte Carlo simulations for the lattice size $L_x=L_y=1000$ with $c=0.4$ (figure \ref{fig10}). As it may be observed in figure \ref{fig10a}, the current does not show the plateau, but instead can be rather well fitted to the expression $j_{\mathrm{x}}(\rho,\alpha)\propto p_{\mathrm{x}} \rho(1-\rho)$ with the factor of proportionality $\approx 0.5756$ close to $1-c=0.6$. In figure \ref{fig10b} is displayed the corresponding histogram of density profile obtained for $\rho=1/2$, which shows a single maximum with a Gaussian-like shape.


\section{Conclusion}

In this work we studied the possibility of avoiding macroscopic phase separation in the exclusion processes with site-wise quenched disorder by allowing particles to bypass defect sites. Bypassing was implemented in a natural way, either by extending the range of hopping or by increasing the dimensionality. In the first example, the performed Monte Carlo simulations indicated the absence of the macroscopic shock provided the hopping length $l$ is taken from the probability distribution obeying the power-law $p_l\sim l^{-(1+\sigma)}$ and $\sigma$ is close to $1$. This was a rather surprising result, since the origin of shocks in one dimension is related to the fact that disorder creates a cluster of slow sites whose length diverges in the thermodynamic limit. However, by studying the entirely segregated model in which all defects are in the consecutive order forming a large ``bottleneck'', we showed that the absence of shocks is a finite-size effect which decays very slowly with the size $l$ of the``bottleneck'' and thus persists for enormously large system sizes, e.g. $l\sim\mathrm{ln}L\approx 50$.

In the second example, we considered the $2d$ exclusion process, which is totally asymmetric in the longitudinal direction and symmetric in the transverse one, with disorder introduced through obstacles in the driving direction. We have shown that a rather simple quantitative description of the process leads to the absence of phase separation in the strongly anisotropic limit, where the hopping rate in the direction of driving is much smaller than in the perpendicular direction, $p_{\mathrm{x}}\ll 2p_{\mathrm{y}}$. In this limit, the original two-dimensional exclusion process reduces to the one-dimensional misanthrope process, in which disorder enters only through the fraction of non-defect sites present in each column of the original $2d$ TASEP. By the central limit theorem, however, the probability distribution of this fraction has a variance which decays as $1/L_{\mathrm{y}}$, so that in the limit $L_{\mathrm{x}}\sim L_{\mathrm{y}}\rightarrow\infty$ the otherwise diverging size of the largest ``bottleneck'' vanishes resulting in the absence of a macroscopic phase separation. On the other hand, if $L_{\mathrm{y}}$ remains finite while $L_{\mathrm{x}}\rightarrow\infty$, as in the multi-lane TASEP with a finite number of lanes, the argument in favour of bottleneck-induced phase separation should be revoked. Obviously, the same argument remains valid for more general types of disorder, provided it obstructs flow only in the direction of driving.

Although the general properties of $2d$ disordered exclusion process remained out of our scope, some interesting open questions emerged. First is the upper value of $p_{\mathrm{x}}$ for which the above mapping applies. Second is the mechanism of phase separation for $p_{\mathrm{x}}$ outside this limit and related to that is the finding of threshold density $\rho_{\mathrm{c}}$. Last, it would be interesting to examine the case where the bond disorder, whose realization may be found in some microfluidic geometries \cite{Champagne10}, is replaced by the site disorder, appropriate for modelling flow of a fluid through porous media or pedestrians avoiding fixed obstacles.

\ack
This work was supported by the Croatian Ministry of Science, Education and Sports through grant No. 035-0000000-3187. 

\appendix
\section*{Appendix}
\setcounter{section}{1}

If $X_i$ denotes the total number of broken bonds in the $i$-th column, $X_i=\sum_{j=1}^{L_{\mathrm{y}}}(1-\omega_{ij})$, then the corresponding probability distribution is the binomial distribution

\begin{equation}
P(X_i=n)=\left({L_{\mathrm{y}} \atop n}\right)c^n (1-c)^{L_{\mathrm{y}}-n},
\end{equation}

\noindent where $c$ is the probability of finding a broken bond at site $(i,j)$. Let $F(x)=P(x<X)$ be the corresponding cumulative distribution and $x^{*}$ its right endpoint, $x^{*}=sup\{x: F(x)<1\}$. We are interested in obtaining the maximum value of $\{X_1,\dots,X_{L_{\mathrm{x}}}\}$ as $L_{\mathrm{x}}\rightarrow\infty$,

\begin{equation}
max\{X_1,\dots,X_{L_{\mathrm{x}}}\}\overset{P}{\rightarrow}x^{*},
\end{equation}

\noindent where $\rightarrow^{P}$ means convergence in probability, since $P(max\{X_1,\dots,X_m\}\leq x)=F^{m}(x)$ is degenerate in the limit $m\rightarrow\infty$ as it converges either to $0$ for $x<x^*$ or to $1$ for $x\geq x^*$. We therefore seek a sequence of positive $a_m$ and real $b_m$ such that $lim_{m\rightarrow\infty}F^{m}(a_m x+b_m)=G(x)$ exists, where $G(x)$ is called the extreme value distribution. A sufficient condition for that is von Mises' condition \cite{extremevalue}, which states that if $F''(x)$ exists and $F'(x)>0$ for $x<x^{*}$ then
 
\begin{equation}
\underset{m\rightarrow\infty}{lim} F^{m}(a_m x+b_m)=\mathrm{exp}\left[-(1+\gamma' x)^{-1/\gamma'}\right],\quad 1+\gamma' x>0,
\label{gx}
\end{equation}

\noindent where $\gamma'$ is given by

\begin{equation}
\underset{t\rightarrow x^{*+}}{lim}\left(\frac{[1-F(t)]F''(t)}{[F'(t)]^2}\right)=-\gamma'-1.
\label{gamma_lim}
\end{equation}

\noindent Moreover, $b_m=U(m)$ and $a_m=mU'(m)$, where $U(m)$ is the inverse function of $1/(1-F(x))$. For $\gamma'=0$, the right-hand side of (\ref{gx}) should read $\mathrm{exp}(-e^{-x})$. To apply this condition, we approximate a binomial distribution with the normal $N(\mu,\sigma^2)$, where $\mu=cL_{\mathrm{y}}$ and $\sigma^2=L_{\mathrm{y}} c(1-c)$. Numerical error in doing so need not worry us since for cumulative distribution it is of the order of $1/\sqrt{L_{\mathrm{y}}}$ (Berry-Essen theorem) and $L_{\mathrm{y}}$ is large. Then it is an easy exercise to show that the cumulative distribution $F(x)$ and its inverse $U(x)$ are given by, respectively,

\begin{equation}
F(x)=\frac{1}{2}\left[1+\mathrm{erf}\left(\frac{x-\mu}{\sqrt{2\sigma^2}}\right)\right],
\end{equation}

\begin{equation}
U(x)=\mu+\sqrt{2\sigma^2}\cdot\mathrm{erf}^{-1}\left(1-\frac{2}{x}\right), \quad x\geq 1,
\end{equation}

\noindent where $\mathrm{erf}(x)$ and ${\mathrm{erf}}^{-1}(x)$ are the error function and its inverse, respectively. Inserting $F(x)$ and its derivatives in (\ref{gamma_lim}) we obtain $\gamma'=0$, i.e. $G(x)=\mathrm{exp}(-e^{-x})$ (Gumbel distribution). The mean and the variance of the Gumbel distribution are given by Euler-Mascheroni constant $\gamma=0.5772\dots$ and $\pi^2/6$, respectively, which gives the mean and the variance of $x^{*}$

\begin{equation}
\langle x^{*}\rangle=a_{L_{\mathrm{x}}}\gamma+b_{L_{\mathrm{x}}},
\end{equation}

\begin{equation}
\langle x^{*2}\rangle-\langle x^{*}\rangle^2=\frac{a_{L_{\mathrm{x}}}^{2}\pi^2}{6},
\end{equation}

\noindent where $a_{L_{\mathrm{x}}}$ and $b_{L_{\mathrm{x}}}$ are given by

\begin{equation}
a_{L_{\mathrm{x}}}(c,L_{\mathrm{y}})=\frac{2\sqrt{2c(1-c)L_{\mathrm{y}}}}{L_{\mathrm{x}}}\cdot \mathrm{exp}\left\{\left[{\mathrm{erf}}^{-1}\left(1-\frac{2}{L_{\mathrm{x}}}\right)\right]^2\right\},
\label{a}
\end{equation}

\begin{equation}
b_{L_{\mathrm{x}}}(c,L_{\mathrm{y}})=cL_{\mathrm{y}}+\sqrt{2c(1-c)L_{\mathrm{y}}}\cdot {\mathrm{erf}}^{-1}\left(1-\frac{2}{L_{\mathrm{x}}}\right).
\label{b}
\end{equation}

\noindent For $L_{\mathrm{x}}\gg 1$, ${\mathrm{erf}}^{-1}(1-2/L_{\mathrm{x}})$ can be expanded around $L_{\mathrm{x}}\rightarrow\infty$, which gives

\begin{equation}
a_{L_{\mathrm{x}}}(c,L_{\mathrm{y}})\approx \left[\frac{4c(1-c)L_{\mathrm{y}}}{2\pi \mathrm{ln} L_{\mathrm{x}}-\pi \mathrm{ln} 2\pi}\right]^{1/2},
\end{equation}

\begin{equation}
b_{L_{\mathrm{x}}}(c,L_{\mathrm{y}})\approx cL_{\mathrm{y}}+\left\{c(1-c)L_{\mathrm{y}}\left[\mathrm{ln}\left(\frac{L_{\mathrm{x}}^{2}}{2\pi}\right)-\mathrm{ln}\mathrm{ln}\left(\frac{L_{\mathrm{x}}^{2}}{2\pi}\right)\right]\right\}^{1/2}.
\end{equation}


\section*{References}

\begin{thebibliography}{99}
    \bibitem{Harris74} Harris A B, {\it Effect of random defects on the critical behaviour of Ising models}, 1974 {\it J. Phys. C: Solid State Phys.}
    {\bf 7} 1671
    
    \bibitem{Stinchcombe02} Stinchcombe R B, {\it Disorder in non-equilibrium models}, 2002 {\it J. Phys.: Condens. Matter} {\bf 14} 1473-87
    
    \bibitem{SchutzDomany93} Sch\"{u}tz G M and Domany E, {\it Phase transitions in an exactly soluble one-dimensional exclusion process}, 1993
    {\it J.   Stat. Phys.} {\bf 72} 277
    \nonum Derrida D, Evans M R, Hakim V Pasquier V, {\it Exact solution of a 1D asymmetric exclusion model using a matrix formulation}, 1993 
    {\it J. Phys. A} {\bf 26} 1493
    
    \bibitem{GwaSpohn92} Gwa L-H and Spohn H, {\it Six-vertex model, roughened surfaces, and an asymmetric spin Hamiltonian}, 1992 
    {\it Phys. Rev. Lett.} {\bf 68} 725–8
    
    \bibitem{Krug91} Krug J, {\it Boundary-induced phase transitions in driven diffusive systems}, 1991 {\it Phys. Rev. Lett.} {\bf 67} 1882
    
    \bibitem{Krug97} Krug J, {\it Origins of scale invariance in growth processes}, 1997 {\it Adv. Phys.} {\bf 46} 139-282
    
    \bibitem{Chowdhury00} Chowdhury D, Santen L and Schadschneider A, {\it Statistical physics of vehicular traffic and some related systems}, 2000
    {\it Phys. Rep.} {\bf 329} 199-329
     
    \bibitem{Chowdhury05} Chowdhury D, Schadschneider A and Nishinari K, {\it Physics of transport and traffic phenomena in biology: from molecular
    motors and cells to organisms}, 2005 {\it Phys. Life Rev.} {\bf 2} 318-52
    
    \bibitem{Derrida98} Derrida D 1998 {\it An exactly solvable non-equilibrium system: the asymmetric simple exclusion process} 
    \textit{Phys. Rep.} \textbf{301} 65-83
    
    \bibitem{Schadscheider10} Schadschneider A, Chowdhury D and Nishinari, K 2010 {\it Stochastic Transport in Complex Systems: From Molecules to
    Vehicles} (Amsterdam: Elsevier)
    
    \bibitem{Mallick96} Mallick K, {\it Shocks in the asymmetry exclusion model with an impurity}, 1996 {\it J. Phys. A: Math. Gen.} {\bf 29} 5375
    
    \bibitem{LeePopkovKim97} Lee H-W, Popkov V and Kim D, {\it Two-way traffic flow: Exactly solvable model of traffic jam}, 1997
    {\it J. Phys. A: Math. Gen.} {\bf 30} 8497
   
    \bibitem{Evans96} Evans M R, {\it Bose-Einstein condensation in disordered exclusion models and relation to traffic flow}, 1996 
    {\it Europhys. Lett.} {\bf 36} 13
    
    \bibitem{KrugFerrari96} Krug J and Ferrari P A, {\it Phase transitions in driven diffusive systems with random rates}, 1996 
    {\it J. Phys. A: Math. Gen.} {\bf 29} L465-71
    
    \bibitem{WolfTang90} Wolf D E and Tang L H, {\it Inhomogeneous growth processes}, 1990 {\it Phys. Rev. Lett.} {\bf 65} 1591
    
    \bibitem{JanowskyLebowitz92} Janowsky S A and Lebowitz J L, {\it Finite-size effects and shock fluctuations in the asymmetric simple-exclusion
    process}, 1992 {\it Phys. Rev. A} {\bf 45} 618
    
    \bibitem{TangLyuksyutov93} Tang L H and Lyuksyutov I F, {\it Directed polymer localization in a disordered medium}, 
    1993 {\it Phys. Rev. Lett.} {\bf 71} 2745
    
    \bibitem{Schutz93} Sch\"{u}tz G M, {\it Generalized Bethe ansatz solution of a one-dimensional asymmetric exclusion process on a ring with
     blockage}, 1993 \textit{J. Stat. Phys.} \textbf{71} 471-505
    
    \bibitem{JanowskyLebowitz94} Janowsky S A and Lebowitz J L, {\it Exact results for the asymmetric simple exclusion process with a blockage}, 
    1994 \textit{J. Stat. Phys.} \textbf{77} 35
    
    \bibitem{Ha03} Ha M, Timonen J and den Nijs M, {\it Queuing transitions in the asymmetric simple exclusion process}, 2003 
    {\it Phys. Rev. E} {\bf 68} 056122
    
    \bibitem{LeeKim09} Lee J H and Kim J M, {\it Directed polymer in random media with a defect}, 2009 {\it Phys. Rev. E} {\bf 79} 051127
    
    \bibitem{TripathyBarma97} Tripathy G and Barma M, {\it Steady state and dynamics of driven diffusive systems with quenched disorder}, 1997 
    {\it Phys. Rev. Lett.} {\bf 78} 3039-42
    
    \bibitem{TripathyBarma98} Tripathy G and Barma M, {\it Driven lattice gases with quenched disorder: Exact results and different macroscopic
    regimes} 1998 {\it Phys. Rev. E.} {\bf 58} 1911–26
    
    \bibitem{Krug00} Krug J, {\it Phase separation in disordered exclusion models}, 2000 {\it Braz. J. Phys.} {\bf 30} 97-104
    
    \bibitem{SzavitsUzelac09} Szavits-Nossan J and Uzelac K, {\it Impurity-induced shocks in the asymmetric exclusion process with long-range hopping},
    2009 {\it J. Stat. Mech.} P12019
    
    \bibitem{SzavitsUzelac08} Szavits-Nossan J and Uzelac K, {\it Totally asymmetric exclusion process with long-range hopping}, 2006 
    \textit{Phys. Rev. E} \textbf{74} 051104
    \nonum Szavits-Nossan J and Uzelac K, {\it Scaling properties of the asymmetric exclusion process with long-range hopping}, 2008
    {\it Phys. Rev. E} {\bf 77} 051116
    
    \bibitem{RamaswamyBarma87} Ramaswamy R and Barma M, {\it Transport in random networks in a field: interacting particles}, 1987 
    {\it J. Phys. A: Math. Gen.} {\bf 20} 2973-87
    
    \bibitem{BarmaRamaswamy93} Barma M and Ramaswamy R, {\it Field-induced transport in random media}, 1993 
    {\it Non-linearity and Breakdown of Soft Condensed Matter} ed K K Bardhan et al (Berlin: Springer) p 309
    
    \bibitem{Saegusa08} Saegusa T, Mashiko T and Nagatani T, {\it Flow overshooting in crossing flow of lattice gas}, 2008 
    {\it Physica A} {\bf 387} 4119-32
    
    \bibitem{Champagne10} Champagne N, Vasseur R, Montourcy A and Bartolo D, {\it Traffic jams and intermittent flows in microfluidic networks}, 2010
    {\it Phys. Rev. Lett.} {\bf 105} 044502
    
    \bibitem{KafriLevine02} Kafri Y, Levine E, Mukamel D, Sch\"{u}tz G M and T\"{o}r\"{o}k J, {\it Criterion for phase separation in one-dimensional
    driven systems}, 2002 {\it Phys. Rev. Lett.} {\bf 89} 035702
    
    \bibitem{AlexanderLebowitz94} Alexander F J and Lebowitz J L, {\it On the drift and diffusion of a rod in a lattice fluid}, 1994 
    {\it J. Phys. A: Math. Gen.} {\bf 27} 683-96
    
    \bibitem{SchmittmannHwangZia92} Schmittmann B, Hwang K and Zia R K P, {\it Onset of spatial structures in biased diffusion of two species}, 1992
    {\it Europhys. Lett.} {\bf 19} (1) 19-25
    
    \bibitem{extremevalue} de Haan L and Ferreira A 2006 {\it Extreme Value Theory: An Introduction} (New York: Springer)
    
    \bibitem{HarrisStinchcombe04} Harris R J and Stinchcombe R B, {\it Disordered asymmetric simple exclusion process: Mean-field treatment}, 2004
    {\it Phys. Rev. E.} {\bf 70} 016108
    
    \bibitem{GordonSchillingWaterman86} Gordon L, Schilling M F and Waterman M S, {\it An extreme value theory for long head runs}, 1986 
    {\it Probab. Th. Rel. Fields} {\bf 72} 279-87
\end{thebibliography}
\end{document}